\documentclass[traditabstract]{aa} 
\usepackage{graphicx}
\usepackage{natbib}
\usepackage{epsfig}
\bibpunct{(}{)}{;}{a}{}{,} 

\usepackage{txfonts}
\begin{document}
\title{Surface composition\thanks{Partially based on observations made with ESO telescopes at the La Silla Paranal Observatory under program ID 286.C-5019(A) and with SOAR telescope. The SOAR telescope is a joint project of Conselho Nacional de Pesquisas Cientificas e Tecnologicas CNPq-Brazil, The University of North Caroline at Chapel Hill Michigan State University, and the National Optical Astronomy Observatory.} and dynamical evolution of two retrograde objects in the outer solar system: 2008 YB$_{3}$ and 2005 VD.}


   \author{N. Pinilla-Alonso
          \inst{1,2} \thanks{The author thanks the Instituto Nacional de Ci\^encia e Tecnologia de Astrof\'isica (INCT-A) for support through a Bolsa a Especialista Visitante (BEV)}\fnmsep\thanks{\emph{Present address:} Department of Earth and Planetary Sciences,1412 Circle Dr, Knoxville TN 37996, USA.}
          \and
	A. Alvarez-Candal
	\inst{1,3}
	\and       
	M. D. Melita
	\inst{4}
 	\and
	V. Lorenzi
	\inst{5}
	\and
	J. Licandro
	\inst{2,6}
	\and
	J. Carvano
	\inst{7}
	\and
	D. Lazzaro
	\inst{7}
	\and
	G. Carraro
	\inst{3}
	\and
	V. Al\'i-Lagoa
	\inst{2,6}
	\and
	E. Costa
	\inst{8}
	\and
	P.H. Hasselmann
	\inst{7}
}
   \institute{Instituto de Astrof\'isica de Andaluc\'ia - CSIC, Apt 3004, 18080 Granada, Spain. \\
   \email{npinilla@seti.org} 
         \and
	Instituto de Astrof\'isica de Canarias -  c/v\'ia L\'actea s/n, 38200 La Laguna, Tenerife, Spain \\
         \and
          European Southern Observatory, Alonso de C\'ordova 3107, Vitacura, Casilla 19001, Santiago 19, Chile \\
          \and
	IAFE (CONICET-UBA), Ciudad Universitaria, Intendente Guiraldes S/N. Buenos Aires. Argentina \\
          \and
          Fundaci\'on Galileo Galilei - INAF, Rambla Jos\'e Ana Fern\'andez P\'erez, 7, 38712 Bre\~na Baja, TF - Spain \\
         \and
         Departamento de Astrof\'isica, Universidad de La Laguna, 38205 La Laguna, Tenerife, Spain \\
	\and
	Observat\'orio Nacional, COAA, Rua Gal. Jos\'e Cristino 77, 20921-400 Rio de Janeiro, Brazil \\           
	\and
         Departamento de Astronom\'ia, Universidad de Chile, Casilla 36-D, Santiago de Chile, Chile \\
             }

   \date{16 July 2012; 28 November 2012}

 
  \abstract   
{Most of the objects in the trans-Neptunian belt (TNb) and related populations move in prograde orbits with low eccentricity and inclination. However, the list of icy minor bodies moving in orbits with an inclination above 40 $^{\circ}$ has increased in recent years. The origin of these bodies, and in particular of those objects in retrograde orbits, is not well determined, and different scenarios are considered, depending on their inclination and perihelion. In this paper, we present new observational and dynamical data of two objects in retrograde orbits, 2008 YB$_{3}$ and 2005 VD. We find that the surface of these extreme objects is depleted of ices and does not contain the `ultra-red' matter typical of some Centaurs. Despite small differences, these objects share common colors and spectral characteristics with the Trojans, comet nuclei, and the group of grey Centaurs. All of these populations are supposed to be covered by a mantle of dust responsible for their reddish- to neutral-color. To investigate if the surface properties and dynamical evolution of these bodies are related, we integrate their orbits for 10$^{8}$ years to the past. We find a remarkable difference in their dynamical evolutions:  2005 VD's evolution is dominated by a Kozai resonance  with planet Jupiter while that of 2008 YB$_{3}$ is dominated by close encounters with planets Jupiter and Saturn. Our models suggest that the immediate site of provenance of 2005 VD is the in the Oort cloud, whereas for 2008 YB$_{3}$ it is in the trans-Neptunian region. Additionally, the study of their residence time shows that 2005 VD has spent a larger lapse of time moving in orbits in the region of the giant planets than 2008 YB$_{3}$. Together with the small differences in color between these two objects, with  2005 VD being more neutral than 2008 YB$_{3}$, this fact suggests that the surface of 2005 VD has suffered a higher degree of processing, probably related to cometary activity episodes.}

 \keywords{Kuiper belt:individual: 2008 YB$_{3}$, 2005 VD - Oort Cloud - Techniques: spectroscopic - Techniques: photometric - Methods: numerical}
\authorrunning{N. Pinilla-Alonso et al.}
\titlerunning{Surface Composition and Dynamical Evolution of two retrograde objects.}
 \maketitle
%

\section{Introduction}

The origin of icy minor bodies in retrograde or high-inclination orbits is puzzling. Only a few objects in the TNb region follow orbits with an inclination larger than 40$^{\circ}$. Among them, 2008 KV$_{42}$, ($a/q/i = 42.3\,\mathrm{AU}/20.33\,\mathrm{AU}/103.4^{\circ}$),the first one discovered, moves in a retrograde orbit \citep{gla09}. \citet{dunlev97} explain that large semi-major axis trans-Neptunian objects (TNOs) with $i$ up to 40$^{\circ}$ might be produced by scattering after gravitational encounters with Neptune during the early evolution of the solar system. But for objects with $i > 40^{\circ}$, the origin could be different. \citet{gla09} discuss different scenarios that could result in the current orbits of these objects. Scattering from the TNb or the Oort cloud after encounters with the giant planets is ruled out based on the low probability of these encounters. Instead, they favor other scenario with an origin in a unobserved reservoir of large-inclination objects beyond Neptune. Recently, \cite{bra12}, studied the origin of the orbits of Centaurs with $q >$ 15 AU and inclinations above 70$^{\circ}$ based on numerical calculations. They show that (1) Centaurs with $q <$ 15 AU can have an origin either in the Oort cloud or in an unobserved reservoir \citep{gla09}; (2) Centaurs with $q >$ 15AU most likely originated in the Oort cloud and were pulled down to their original orbit after encounters with Neptune and Uranus. 

On the other hand, there is a population of minor bodies in the outer solar system among which high-inclination orbits are dominant, the Damocloids. \citet{jew05} introduced this term to refer to objects with orbits like those of the Halley-type comets (HTC), but without signs of outgassing. They would be the dormant nuclei of the HTC. Their distribution of inclinations is very similar to that of the long-period comets, with most of them having large inclinations and several moving in retrograde orbits. For simplicity, he defined this group of objects based on their Tisserand parameter respective to Jupiter\footnote{The Tisserand parameter is a constant of the motion in the restricted circular three-body problem. In particular T$_{J}$ provides a simple estimation of the gravitational influence of Jupiter in the orbit of other objects in the solar system}. As it happens for the HTC, Damocloids will have T$_{J} <$2. As \citet{jew05} noted, in spite of their  wide range of orbital eccentricities and inclinations, Damocloids share a common characteristic, namely the high value of these two orbital parameters.

In this work, we study two retrograde objects with perihelion smaller than 7 AU, 2008 YB$_{3}$ ($a/q/i=11.65\,\mathrm{AU}/5.01\,\mathrm{AU}/172.901^{\circ}$) and 2005 VD ($a/q/i=6.68\,\mathrm{AU}/6.49\,\mathrm{AU}/105.035^{\circ}$). According to the Minor Planet Center (MPC) definition of Centaur, these objects are Centaurs in retrograde orbits; however, according to the value of their T$_{J}$  (-1.23 and 0.14, respectively), they are Damocloids. In Section ~\ref{sec:data}, we present new visible colors of 2008 YB$_{3}$ and the first colors of 2005 VD. We also show the first spectra of 2008 YB$_{3}$, from 0.35 to 2.3  $ \mu$m. In Section ~\ref{sec:analysis}, we compare these to the colors and spectra of other known populations of small bodies, and in Section ~\ref{sec:models}, we discuss how the dynamical history of these objects can be related to their present surface characteristics. We discuss and summarize the results in Section ~\ref{sec:conclusions}.


\section{Observations and data reduction}

\label{sec:data}
We present data of the two retrograde Centaurs 2005 VD and 2008 YB$_3$. We obtained photometric measurements of both, at different telescopes, while spectroscopy was only obtained for the latter at two different telescopes and epochs: before the perihelion passage at the Telescopio Nazionale Galileo (TNG), and after the passage at the ESO-Very Large Telescope. Below we describe in detail the observations.

\subsection{2005 VD}

Photometric observations of 2005 VD using the Bessel-BVR filters were acquired with the SOAR Optical Imager (SOI), mounted on the 4-m SOAR telescope on Cerro Pachon, Chile. The observations were made during a service mode run on April 3, 2011. The night was photometric, with a median seeing below $1$". At the time of the observation, 2005 VD was at heliocentric and geocentric distances of, respectively, 7.44 and 6.60 AU, and at a phase angle of $4.5^{\rm o}$. The observations were made  using $2\times2$ binning, non-sideral guiding, and exposures times of  940, 1200, and 1480 seconds for the B, V, and R filters, respectively. The object was observed  at two instants during the night, at airmass around 1.2 and 1.08. To correct for extinction, we used several stars in the field of the photometric Landolt standard star PG1323, which were observed at three different airmasses during that night  with the same instrumental setup as the asteroid.

The data reduction consisted of overscan, bias, and flat field corrections for all filters. The instrumental magnitudes of the asteroid and of the star were obtained through aperture photometry,  and the observations of the star were then used to derive the extinction coefficients and zero points of the night for each filter. The resulting calibrated magnitudes and averaged color indexes are shown in Table \ref{tab:colors}.

\subsection{2008 YB$_3$}

\subsubsection{Photometry}

{\bf Photometry in the visible domain}

\begin{table*}

	\begin{center}
 	\caption{Color indexes of the retrograde objects observed.}
 	\label{tab:colors}
 	\begin{tabular}{lcccccll} 
 		\hline	
		Object  & H &  V & (B-V) & (V-R) & (V-I) &Observatory& Reference\\
		\hline
		2005 VD & 14.2 & $ 23.45\pm0.09 $ & $0.60\pm0.17$  & $0.45\pm0.11$ &    &SOAR&this work\\
		2008 YB$_3$&9.5 & $18.231\pm0.001$ & $0.82\pm0.01$\tablefootmark{a} & $0.50\pm0.06$ & $1.00\pm0.08$ &CTIO&this work\\
		2008 YB$_3$&& $$ & $0.82\pm0.01$ & $0.46\pm0.01$ & $1.29\pm0.01$ &&Sheppard (2010)\\
		Sun&&&0.64&0.36&0.69&&Hainut \& Delsanti (2002)\\
		\hline
		
 	\end{tabular} 
	
	\tablefoottext{a} from \cite{she10}\\
	\end{center}
\end{table*}

We observed 2008 YB$_3$ with the Y4KCAM camera attached to the Cerro Tololo Inter-American Observatory (CTIO) 1-m telescope, 
operated by the Small \& Moderate Aperture Research Telescope System consortium\footnote{\tt http://www.astro.yale.edu/smarts} (SMARTS) during three consecutive nights in December 28-30, 2010. All observations were carried out in photometric, good-seeing (always less than 1.2$"$), conditions. About 40 exposures per night were 
taken in V,R, and I filters, with typical exposures of 200 s. 

Basic calibration of the CCD frames was done using the IRAF\footnote{IRAF is distributed by the National Optical Astronomy 
Observatory, which is operated by the Association of Universities for Research in Astronomy, Inc., under cooperative 
agreement with the National Science Foundation.} package CCDRED. For this purpose, zero exposure frames and twilight sky 
flats were taken every night.

Because the object was transiting a crowded region of the sky, about 20$^{\rm o}$ away from the plane of the galaxy, we used aperture correction techniques to extract the magnitudes. We chose a few isolated stars, typically 20, to compute the correction using the method of curve of growth \citep{ste90}. To determine the transformation from the instrumental system to the standard Johnson-Kron-Cousins system and to correct for extinction, we observed stars in Landolt's areas SA 98 \citep{lan92} multiple times and with different airmasses ranging from $\sim1.03$ to $\sim2.0$ and covering quite a large color range -0.3 $\leq (B-V) \leq$ 1.7 mag. The calibrated colors are shown in Table~\ref{tab:colors}. We note that these colors were obtained as the average over several measurements taken during the run.

\begin{figure}[ht]
 \includegraphics[width=8.8cm]{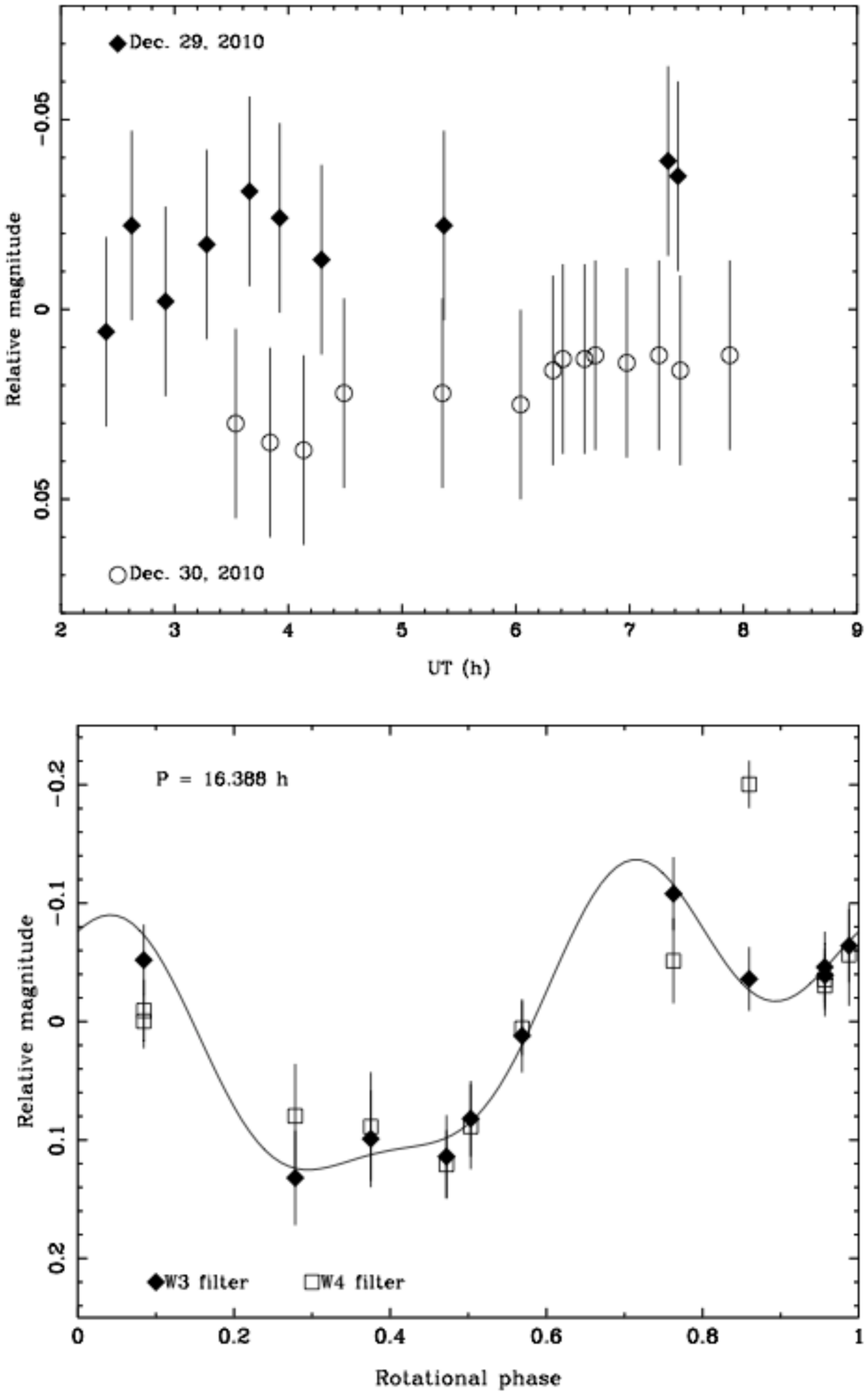}
 \caption{Lightcurves of 2008 YB$_{3}$. Above lightcurve obtained with the CTIO 1-m telescope operated by the SMARTS consortium. Below, lightcurve obtained with the WISE is shown. We estimate a rotational period based on this lightcurve of  16.388 $\pm$0.002 hr. }
 \label{fig:lc}
\end{figure}

Since we wanted to obtain the photometric lightcurve, we used as base the R filter observations, with alternating observations in the V and I filters. The lightcurves were obtained using differential magnitudes: $\Delta m = m_{\rm ast} - m_{\rm sta}$. Out of the three nights, the first one could not be used for lightcurve analysis due to the scatter in the relative magnitude product of the object passing close to stars in the background. The individual lightcurves obtained from the useful data are shown in Fig.~\ref{fig:lc}.\\

\noindent
{\bf Photometry in the mid-infrared domain}

The Wide-field Infrared Survey Explorer (WISE) surveyed the entire sky in four infrared wavelengths (3.4, 4.6, 12.0, and 22.0 $ \mu$m, denoted W1, W2, W3, and W4, respectively) during its 2010 mission \citep{wri10}. Following the procedure described in \citet{mai11}, we queried the WISE Preliminary Release Single Exposure (L1b) Working Database of the Infrared Science Archive\footnote{\texttt http://irsa.ipac.caltech.edu/Missions/wise.html}. This search returned WISE data of 2008 YB$_{3}$, obtained on April 10-11, 2010. As recommended by \citet{mai11}, we discarded data that did not meet a number of requirements according to their reported magnitude uncertainties and quality and artifact flags. Since most of the data in bands W1 and W2 were rejected, only W3 and W4 magnitudes have been used. Given that we are interested in relative differences within each one of the lightcurves, which are constructed separately for each band, none of the further corrections suggested in \citet{wri10} and \citet{mai11} were applied. 

\subsubsection{Spectroscopy}

{\bf TNG: Spectra obtained before the pass by the perihelion.}

\begin{table*}
 	\caption{Observational parameters of the spectroscopic observations of 2008 YB$_{3}$. See text for details. }
 	\label{tab:spectra}
 	\begin{tabular}{lccccccccc}
	\hline
Instrument & Date & UT & Airmass & r & $\delta$ & $\alpha$ & m$_{V}$ & n$\times$T$_{exp}$ & Solar\\
& &start & & (AU) & (AU) & ($\circ$) & &(secs) & Analog\\
\hline
LRS-LRB&2011.01.08& 2:25&1.44&6.49&5.71&5.7&17.8&$3\times600$&L98-978, L102-1081, L93-101\\
LRS-LRR&id.& 2:57&1.48&id.&id.&id.&id.&$2\times900$&id.\\
NICS Amici &id.& 5:13&2.51-2.98&id.&id.&id.&id.&$16\times90$&L102-1081, L107-684, L107-998\\
X-shooter(UVB) & 2011.01.28 & 2:52 & 1.02-1.05 &6.49&5.67&5.2&17.8& $4\times1100$ & L98-978\\
X-shooter(VIS) & id. &id.&id.&id.&id.&id.&id.& $4\times1140$ & id.\\
X-shooter(NIR) & id. &id.&id.&id.&id.&id.&id.& $4\times1200$ &id.\\
\hline
\end{tabular}
 \tablefoot{For the observation with X-shooter the three different exposure times correspond to the same observation, so all the other parameters
 are identical. For the TNG, observations in LRB, LRR and Amici have been done on the same date but on different expositions.}

\end{table*}

Visible spectra of 2008 YB$_{3}$  were done with the 3.58m TNG (El Roque de los Muchachos Observatory, Canary Islands, Spain) on January 8, 2011. The spectrograph DOLORES with the LR-R and LR-B grism and the 2.0"  slit width was used. With the grism LR-B we obtained two spectra with exposure times of 900 sec, covering the  0.35  to 0.7   $ \mu$m spectral range (see Table~\ref{tab:spectra}). In LR-R we obtained three spectra of 600 sec, covering the 0.45 to 1.0   $ \mu$m range. The object was shifted in the slit by 10$"$ between consecutive spectra to better correct the fringing. We did not dither in the slit when using LR-B because fringing does not affect the blue grism. The spectra obtained with each grism were averaged and then joined using the overlapping wavelength range. 

\begin{figure}
   \centering
    \includegraphics[width=9.2cm]{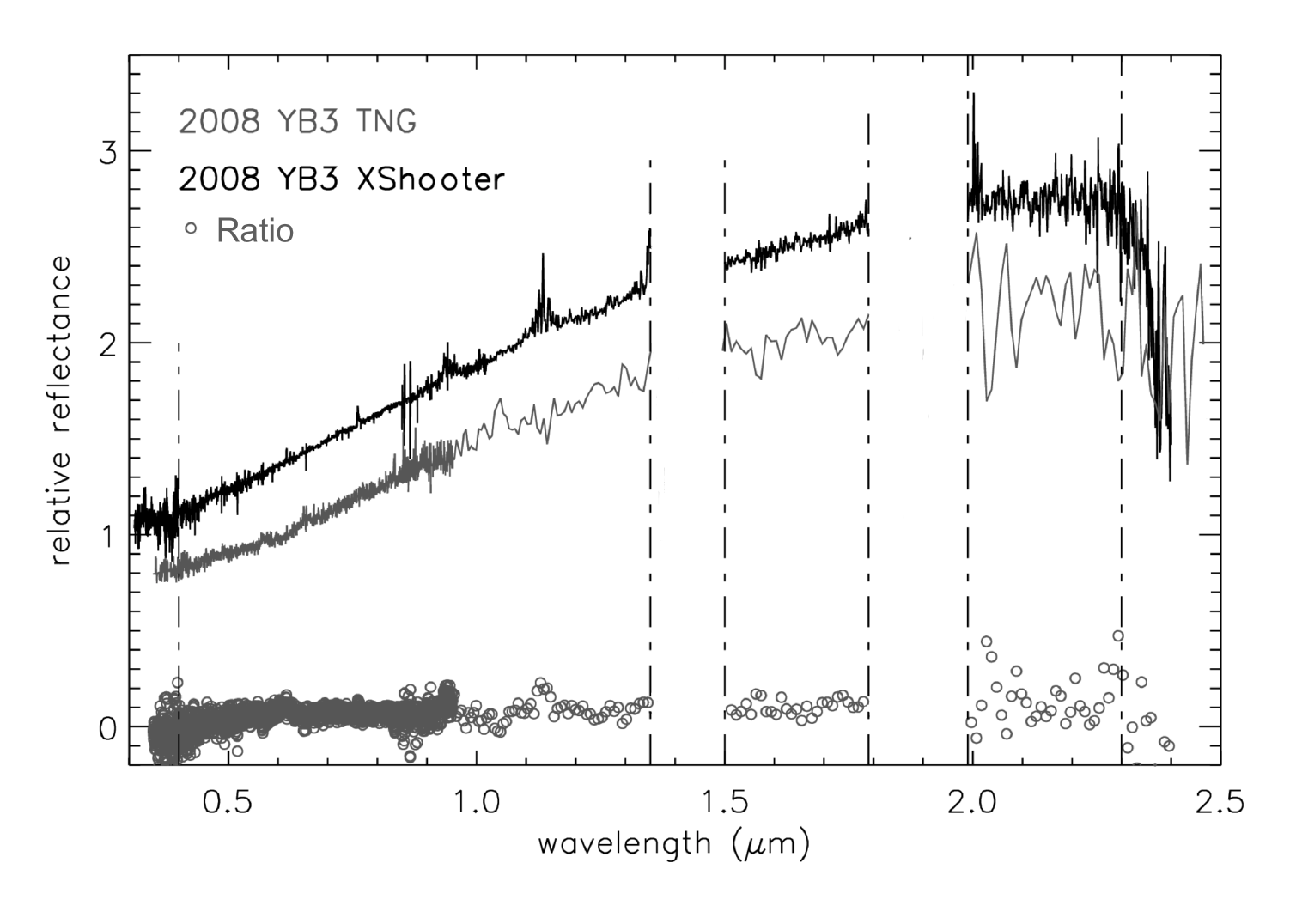}
    \caption{Spectrum of 2008 YB$_{3}$ from X-shooter and TNG. Both are normalized to unity at 0.6   $ \mu$m. The spectrum from X-shooter has been shifted 0.3 for clarity. The flat {\it spectrum} shown at the bottom is the ratio between the two spectra shown in the figure.}
     \label{fig:spectra}
\end{figure}

To correct for telluric absorption and to obtain the relative reflectance, the G stars Landolt (SA) 98-978, Landolt (SA) 102-1081, and Landolt (SA) 93-101 \citep{lan92} were observed at different airmass (similar to those of the Centaur), before and after the Centaur observations, and used as solar analog stars. The spectrum of each object was divided by those of the solar analog stars observed the same night and at similar airmasses and then normalized to unity around 0.55  $ \mu$m, thus obtaining the normalized reflectance. The obtained spectrum, after merging with the near-infrared (NIR) obtained on the same night as explained below, is shown in Fig.~\ref{fig:spectra}.

The NIR spectrum was obtained using the high throughput, low-resolution spectroscopic mode of the Near-Infrared Camera and Spectrometer (NICS) at the TNG,  with an amici prism disperser. This mode yields a complete 0.8-2.5  $ \mu$m spectrum. We used a 1$"$ wide slit corresponding to a spectral resolving  power R$\sim$50, quasi constant along the spectrum. The slit was oriented at the parallactic angle, and the tracking was at the Centaur proper motion. The acquisition consisted of a series of 90-second exposure times in one slit position (position {\em A}), and then the telescope was offset by $10''$ in the direction of the slit (position {\em B}). This process was repeated, obtaining four {\em ABBA} cycles up to a total exposure time of 1440 seconds. We used the observing and reduction procedure described by \citet{licreduc}.

To correct for telluric absorption and to obtain the relative reflectance, the G2 stars Landolt (SA) 102-1081, Landolt (SA) 107-684, and Landolt (SA) 107-998 \citep{lan92} were observed at different airmasses during the night, before and after the Centaur observations, and used as solar analog stars.
The spectrum of 2008 YB$_{3}$, obtained by averaging all the individual AB pairs, was divided by the spectrum of each of the solar analog stars. Finally, all the different spectra calibrated in wavelength were averaged and the resulting spectrum was normalized to unity around 1.0  $ \mu$m, thus obtaining the scaled reflectance. In Fig.~\ref{fig:spectra}, the resulting spectra obtained by combining all the extracted $AB$ of each night and normalized to join the visible spectrum around 0.9  $ \mu$m are presented.

Around the two large telluric water band absorptions, the signal-to-noise ratio (SNR) of the spectrum is very low. Even in a rather stable atmosphere, the telluric absorption can vary between the object and solar analog observations, introducing false features. Therefore, any spectral structure in the 1.35-1.46 and 1.82-1.96  $ \mu$m regions can produce false features. There are also a few telluric absorption regions that are not as deep, for example, 0.93-0.97, 1.10-1.16, 1.99-2.02, and 2.05-2.07  $ \mu$m. Features in these regions must be carefully checked.\\

\noindent
{\bf Very Large Telescope (VLT): spectra obtained after the pass by the perihelion}

Spectroscopy from the visible up to the NIR was obtained with X-shooter at the 
ESO-VLT, unit 2 Kueyen. X-shooter is an {\it Echelle} spectrograph able to obtain
the spectrum between 0.3 and 2.48   $ \mu$m in one shot \citep{dod06}.
We observed 2008 YB$_3$ on the night of January 28, 2011 under good conditions, using the slit
mode of X-shooter. We used the $2\times1$ binning for the UVB and VIS detectors with 
slit widths of 1.0\arcsec, 0.9\arcsec, and 0.9\arcsec, for the UVB, VIS, and NIR 
arms, respectively, giving a resolving power of about 5,000 per arm. 

To remove the solar and telluric signals from the spectra, 
we observed the star SA98-978 \citep{lan92} with the same observational setup as
2008 YB$_3$ and a similar airmass to minimize effects of differential refraction.
Details of the observations are presented in Table~\ref{tab:spectra}.
We note that the exposure times are different for the three arms. They were selected
to minimize deadtime due to readout, especially in the UVB and VIS arms.

The data were reduced and treated as in \citet{alv11}, but using the version 1.2.2 of the X-shooter pipeline.
Implementing this tool, the data were bias- and dark-subtracted, flat-fielded, wavelength-calibrated,
corrected by atmospheric extinction, and sky-subtracted. The extraction of the spectra 
was made from a two-dimensional merged file, generated by the pipeline, using IRAF.
Once all spectra were extracted, we divided those of 2008 YB$_3$ by the star, which was used 
as a telluric and solar analog star, and cleaned the spectra of remaining bad pixels. Since the resulting spectrum
(Fig. \ref{fig:spectra}) is featureless, we applied a re-binning of 5 pixels, which resulted in a spectral resolution of 1000.
Some artifacts introduced by the star and poor telluric correction could be seen around 0.95 and 1.15 $ \mu$m.

\section{Analysis of the data}
\label{sec:analysis} 

One of the goals of the visible photometric observations was to study 2008 YB$_{3}$ rotational properties. Unfortunately, the object was crossing a crowed region of the sky and many datapoints had to be discarded due to contamination from nearby stars. 

The top panel of Fig.~\ref{fig:lc} shows the two individual lightcurves in the visible, which show a slow increase in magnitude of about 0.02. The total time span in each observation was about 4 h. We will not give any constraint on the rotational period based on these data, given that the apparent variation in brightness is of the order of the error bars.

The WISE data have a much larger time coverage, about 16 h in two filters, and they are better suited to search for a rotational period. We used the method described in \citet{harlup89}, fitting a Fourier polynomial of 3rd order to the W3 filter data. The fit also works for the W4 data if we ignore one point at about 14 hr that is not very reliable. The fitted data phased to this period are shown in Fig.~\ref{fig:lc}. The rotational period we obtained is (16.388$\pm$0.002) h. We note that this is a rough estimation of the period based on a sparse coverage of the lightcurve, so the error is probably underestimated.

The visible and infrared lightcurves cannot be phased together due to the time span between them (almost 3 months) and the different wavelength. One possible explanation for the huge difference between the two is a change in the aspect angle leading to an almost polar view during the visible observations.

As mentioned in Section~\ref{sec:data}, the photometric data, for 2005 VD and 2008 YB$_{3}$, were averaged to extract color indexes and to be compared with the data of other populations of minor bodies. The averaged visible colors for both Centaurs are shown in Table.~\ref{tab:colors}. For 2008 YB$_{3}$, our colors are very close to those published by \citet{she10}. For 2005 VD, we obtain neutral colors. Due to the faintness of this object, this is the only color determination for 2005 VD available in the literature.

\begin{figure}[t]
 \includegraphics[width=8.8cm]{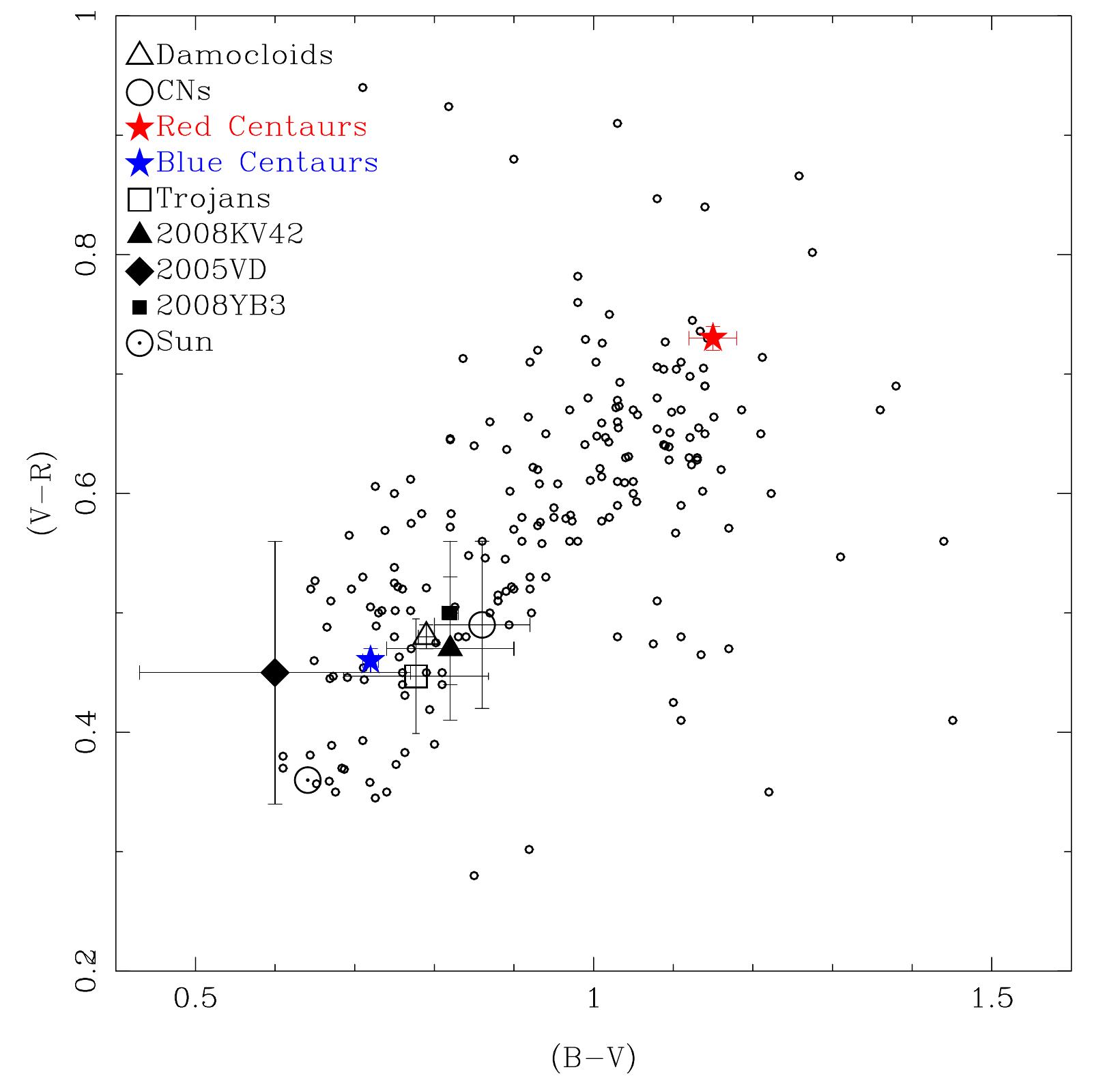}
 \caption{ (B-V) vs (V-R) color indexes of 2008 YB$_{3}$ (black square) and 2005 VD (black diamond). We include colors of some objects in the primitive minor bodies for comparison. See text for references. Background black points are TNOs from \citet{ful08}. Empty diamond is the average color of Damocloids from \citet{jew05}. Empty circle is the average color of nuclei of Jupiter Family comets from \cite{lamtot09}. Red and blue squares represent the red and blue groups identified by \citet{mellic12}. Asterism represents colors of Trojans from \citet{eme11}. In this representation we do not distinguish between the two groups identified by \citet{eme11} because they are not distinguishable in the visible wavelength. We include 2008 KV$_{42}$, a retrograde TNO, from \citet{she10}.
}
 \label{fig:colors}
\end{figure}

\begin{figure}[t]
 \includegraphics[width=9.2cm]{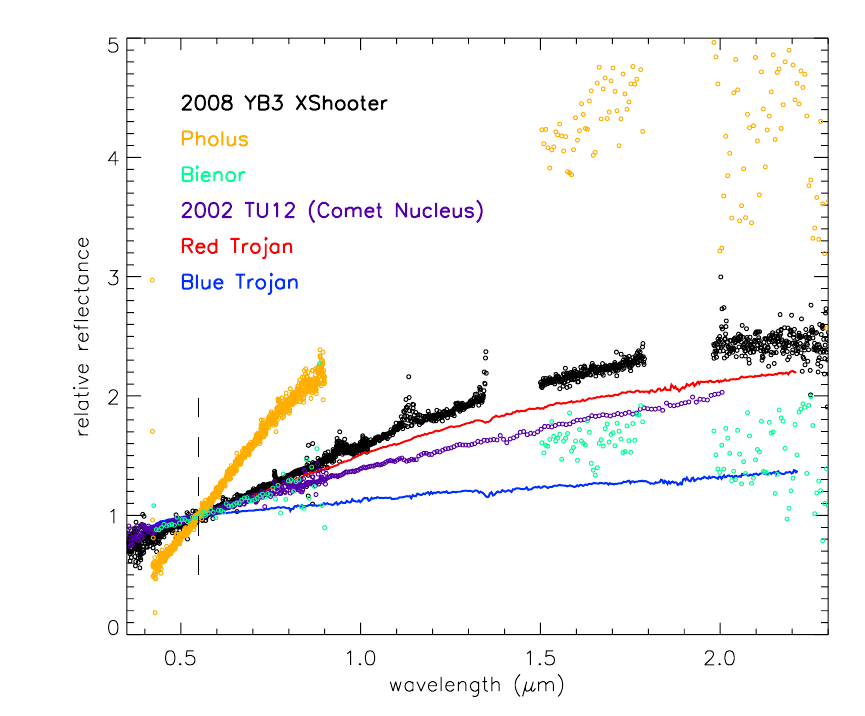}
 \caption{We show the spectrum of 2008 YB$_{3}$ obtained with X-shooter, and some other primitive minor bodies, for comparison: a red Centaur, Pholus \citep{bar11,per10,for09,ful08}; a grey Centaur, Bienor \citep{dem09,gui09,alv08}; the average spectrum of the red and grey Trojans \citep{eme11}; the spectrum of 2004 TU$_{12}$, a red comet nuclei \citep{cam06}. The retrograde Centaur is very similar but redder than all the other primitive bodies, with the exception of Pholus, where ultra-red matter and ices have been detected.}
 \label{fig:spcomp}
\end{figure}

The slightly red color of 2008 YB$_3$ is confirmed by the characteristics of its spectrum as seen in Fig.~\ref{fig:spectra}.
This image shows two spectra taken within 20 days. The two are similar throughout the complete wavelength range, as shown by the small residues resulting from the calculated ratio of both spectra, which are included in the same figure. It is interesting to note that one spectrum was taken before the perihelion passage (at $\sim$6.5 AU), while the other was taken after it. We did not find any report of activity for this object in the literature. We checked carefully all the images taken during our observations and did not find any traces of activity. This suggests that the actual inventory of ices on the surface of this Centaur, or close to it, is not enough to trigger an active episode at its distance to the Sun.

We study in detail the spectrum from X-shooter because it has a higher SNR and resolution. We do not report any absorption feature in the whole range. The main
characteristic of this spectrum is a reddish slope that changes along the spectral range covered. \citet{she10} computed a normalized spectral gradient for the optical colors of 9.6 $\pm$ 0.5. We calculated the reflectivity gradient $S'[\%/1000$ \AA$]$ over three different ranges where the spectrum can be fitted by a straight line. We found that the slope decreases from the near-ultraviolet to the NIR domains, as these values show $S'_{1}$=12.86 $\pm$ 0.07 (from 0.4 to 1.35  $ \mu$m), $S'_{2}$=7.41 $\pm$ 0.01 (from 1.4 to 1.79  $ \mu$m), and $S'_{3}$= 1.32 $\pm$ 0.15 (from 1.99 to 2.3  $ \mu$m). The fit of the spectrum by a straight line is sensitive to the first and last points of the interval considered and is used to estimate the error in the slope. It is worth bearing in mind that the X-shooter spectrum is obtained {\it simultaneously} in three different arms. Beacuse there is a small overlap between the spectral ranges covered by each arm, mounting the complete spectrum is trivial and does not rely on any further observation, such as photometric data. There is a significant absorption on the spectrum above 2.3 $ \mu$m that was not considered for the calculation of the slopes because this part of the spectrum could be degraded by the sensitivity decay of the spectrograph.

The absence of clear features in the spectrum of 2008 YB$_3$ and of colors of 2005 VD in the NIR prevents us from obtaining further knowledge of the surface composition of these objects however some clues can be gained by comparing colors and spectrum with other minor objects in the outer solar system. We are particularly interested in the comparison with other primitive minor bodies with different dynamical or physical evolution: Centaurs, Trojans, Damocloids, and nuclei of Jupiter Family Comets. In this comparison we also consider, when data are available, the other retrograde object discussed in \citet{gla09}: the TNO 2008
KV$_{42}$. 

The colors of the Centaurs follow a clear bimodal distribution that has been object of study for years \citep{pei03}. We represent these two groups in Fig.~\ref{fig:colors} by the average of the colors of the groups studied in \cite{mellic12}. In this work, the authors explain that this bi-modality could be associated to a different degree of processing of the surface. Centaurs in the red group have surfaces with a greater amount of ices,  whereas the surfaces of the Centaurs in the  grey group are covered by  dust mantles that hide the ices. Dynamical studies show how all the objects in the second group spend a larger part of their life in orbits that bring them closer to the Sun, which implies a higher degree of processing of the surface. 

Another interesting group to include in the comparison is the Trojan population. Recently, \cite{eme11} identified two clear groups when considering the slope in the NIR. One, the redder group, is similar to the traditional D-type asteroids, whereas the less-red group is more similar to the X-type asteroids. The authors suggest these differences are best explained as differences in intrinsic composition, related to a different origin, rather than an effect of surface modification. Therefore, the red group would be compatible with an origin in the domain of the volatile ices, while the less-red group would be compatible with capture by Jupiter from near circular orbits at $\sim$ 5 - 6AU. 

The third group we consider for comparison is the nuclei of Jupiter Family Comets. The aspect of these objects is determined by successive epochs of activity that result in surfaces covered by dust mantels that exhaust or mask the ices. The colors of the comet nuclei are neutral to reddish \citep{lamtot09}, and their spectra are featureless. In particular, we show in Fig.~\ref{fig:spcomp} the reddest spectrum obtained from a cometary nucleus, 162P/Siding Spring (2004 TU$_{12}$, \citeauthor{cam06} \citeyear{cam06}).

Finally, we include Damocloids in the comparison. The origin of Damocloids is still a matter of discussion. These objects have been defined in \citet{jew05} as the nucleus of Halley Comets, which have a probable source in the Oort cloud \citep{emebai98}. Dynamical theories show that the origin of these objects could be far from the TNO population. In fact, these objects were probably formed closer to the Sun than they are now and then scattered by Jupiter to high-inclination orbits with a large semi-major axis in the very early epoch of formation of the solar system. Unfortunately, we could not find any spectrum of a Damocloid in the visible and NIR, so we consider colors for comparison as can be seen in Fig.~\ref{fig:colors}.

From the comparison of colors and spectra, we find that the retrograde objects have neutral- to reddish- colors that indicate absence of ultra-red material, which is present on Centaurs in the red lobe. The colors of the retrograde objects are more similar to those of surfaces with low content of ices, as comet nuclei, Trojans, Damocloids, and the grey group of Centaurs.There is some difference between the colors of 2005 VD and the colors 2008 YB$_{3}$, the former being more neutral than the latter. This difference is well within the (very large) error bars and suggests that the dust mantle covering the surface of 2005 VD could be more processed as a result of  a larger residence of 2005 VD at shorter perihelion distance. To check if this reasoning is supported by the dynamical history of both objects we studied their orbits and how they have evolved from the Oort cloud to their present position.

\section{Dynamical Models} \label{sec:models} 
In this section, we investigate the dynamical evolution of the orbits of 2005 VD and 2008 YB$_{3}$.  At the present time, the orbits of these objects have their perihelion in the region of the giant planets. Some theories \citep{ferbru00, don04, dun87} concur in describing  the formation site of the icy minor bodies in the Oort cloud and in the TNb.
However, an extrasolar origin of inner cloud comets has also been put forward \citep{lev10,bra06}. We assume that these reservoirs are the most plausible immediate sites of provenance of the retrograde objects treated here, as they have plausible site-of-origin relationships with the Halley-type comets \citep{gla09} and the consideration of an extrasolar formation site is beyond the scope of this investigation. The orbits of Centaurs are unstable, with a typical lifetime of the order of 10$^{7}$ years. Therefore we model their motion for 10$^{8}$ years.

We have considered the objects as point-masses under the gravitational field of the Sun and the four major planets. We obtained the initial conditions, in the form of osculating orbital elements for the planets and the minor bodies from the JPL Horizons website (http://ssd.jpl.nasa.gov/?ephemerides), corresponding to Julian date
$2454101.5$ (see Table~\ref{tab:dyn}). The equations of motion were integrated implementing the numerical integrator scheme used in \cite{brumel02}, which is a hybrid symplectic second-order method that treats close encounters with the six planets using a Burlish and Stoer integrator with the strategy developed by \citet{cha99}.  We set
the integration time-step of the symplectic method equal to $0.01$yr. The integrations were conducted into the past and stopped when the semi-major axis exceeds a value of $20,000$ AU or the heliocentric distance is smaller than $0.1$ AU or bigger than $100,000$ AU.

\begin{table}[ht]
\caption{JPL orbital elements of 2005 VD and 2008 YB$_{3}$ for Julian date $2454101.5$.}
\label{tab:dyn}
\begin{center}
	   \begin{tabular}{|l|c|c|} \hline
	   Orbital element (unit)   & 2008 YB$_{3}$  & 2005 VD\\
	   \hline
	   Semi-major axis (AU) & 11.6573444  & 6.6652109  \\
	   Eccentricity & 0.4435901 &  0.2488274\\
	   Inclination ($\degr$) & 105.05031 &  172.91172 \\
	   Longitude of node ($\degr$) & 330.51412 & 178.4653 \\
	   Argument of perihelion ($\degr$) & 112.47615  & 173.03421 \\
	   Mean anomaly ($\degr$) & 354.51704 & 90.76943 \\
	   \hline
	   \end{tabular}
\end{center}
\end{table}

The orbital elements used as initial conditions of our calculations are given in Table~\ref{tab:dyn}. When the initial conditions of the orbits given in Table~\ref{tab:dyn} are integrated into the past, we noticed that the dynamical evolution of both objects is qualitatively different. The total time of integration of 2008 YB$_{3}$ was $-20.4887$ Myr and $-6.3891$ Myr for 2005 VD.

The inclination of orbit of 2005 VD experiences the largest variations between $120\degr$ and $180\degr$. In particular an increase about t=$5\times10^5$yr corresponds to a time-interval in which the longitude of perihelion librates about $0\degr$. At this time, there is also a decrease in eccentricity. This is evidence that the orbit of this object has evolved into a Kozai resonance \citep{koz62}. The perihelion penetrates into the region of the terrestrial planets and is below the semi-major axis of Jupiter for most of the lifetime span. The aphelion distance remains bounded between $300$AU and $10$AU up to $t \approx 1.5$Myr. At this point, the longitude or perihelion starts
circulating and the aphelion distance penetrates into trans-Neptunian distances, where it is found at present. The perihelion distance approximately ranges between $2$ AU and $6$ AU, until it is expelled after a very energetic close encounter with planet Jupiter. Such event occurred at approximately $5$Myr, at a relative distance of $46$
planetary radii. Most of the dynamics of this object is dominated by encounters with planet Jupiter, with the perihelion distance crossing in and out the orbit of the planet several times in its lifetime.

On the other hand, the orbit of 2008 YB$_{3}$ remains quasi-polar, with inclination values ranging between $100\degr$ and $110\degr$ for the whole dynamical lifetime span. The aphelion distance evolves smoothly from $\approx 1000$AU to its present value.  The longitude of perihelion oscillates with decreasing frequency as the object is
transported outwards. Most of the dynamics is dominated by close alternate encounters with planets Jupiter and Saturn, as the perihelion distance is bounded by their orbits. Repeated close encounters finally produce the expulsion of the object.

Naturally, the orbital elements are known within a specific uncertainty interval. Therefore, the orbital evolution of these objects may not be described by the single numerical integration of their nominal orbital elements. To explore the full set of initial conditions that correspond to the presently observed orbits of these bodies, we integrated the orbits of $100$ {\it clones} for each asteroid. The initial orbital elements ${\bf a}$ of each clone $i$, ${\bf a}_i$, were calculated as \[ {\bf a}_i = {\bf a}_0 + {\bf \sigma({\bf a})} \times N(0,1), \] where ${\bf a}_0$ are the nominal orbital elements used in the previous experiments, $N(0,1)$ is a Normal distribution with mean equal to $0$ and standard deviation equal to $1$, and $\sigma({\bf a})$ is the corresponding $1 \sigma$ deviation of each element, obtained from the AstDyS database (http://hamilton.dm.unipi.it/astdys) for Julian Date $2454101.5$. These values are given in Table~\ref{tab:sigmas}.

The integrations were stopped for particles reaching the criteria given above. The calculations were followed for a timescale of $10^8$
years in the past, when only one clone corresponding to object 2008 YB$_{3}$ remained and none corresponding to 2005 VD.

\begin{table}[ht]

	\caption{1-sigma standard deviations of the orbital elements of the orbits of 2008 YB$_{3}$ and 2005 VD, used to define the orbital elements of the orbits of the clones.}
	\label{tab:sigmas}
	\begin{center}
	\begin{tabular}{|l|c|c|} \hline
	Orbital element (unit)   & 2008 YB$_{3}$  & 2005 VD\\
	\hline
	Semimajor axis (AU) & 0.000471  & 0.001605  \\
	Eccentricity & 0.00002077 & 0.0001479 \\
	Inclination ($\degr$) & 0.0000549 & 0.0003509 \\
	Longitude of Node ($\degr$) & 0.0001475 & 0 .003448 \\
	Argument of perihelion ($\degr$) & 0.002243 & 0.03649 \\
	Mean Anomaly ($\degr$) & 0.0008794 & 0.01661 \\
	\hline
	\end{tabular}
	\end{center}
\end{table}

\begin{figure*}
  \centering
  \includegraphics[width=0.49\linewidth]{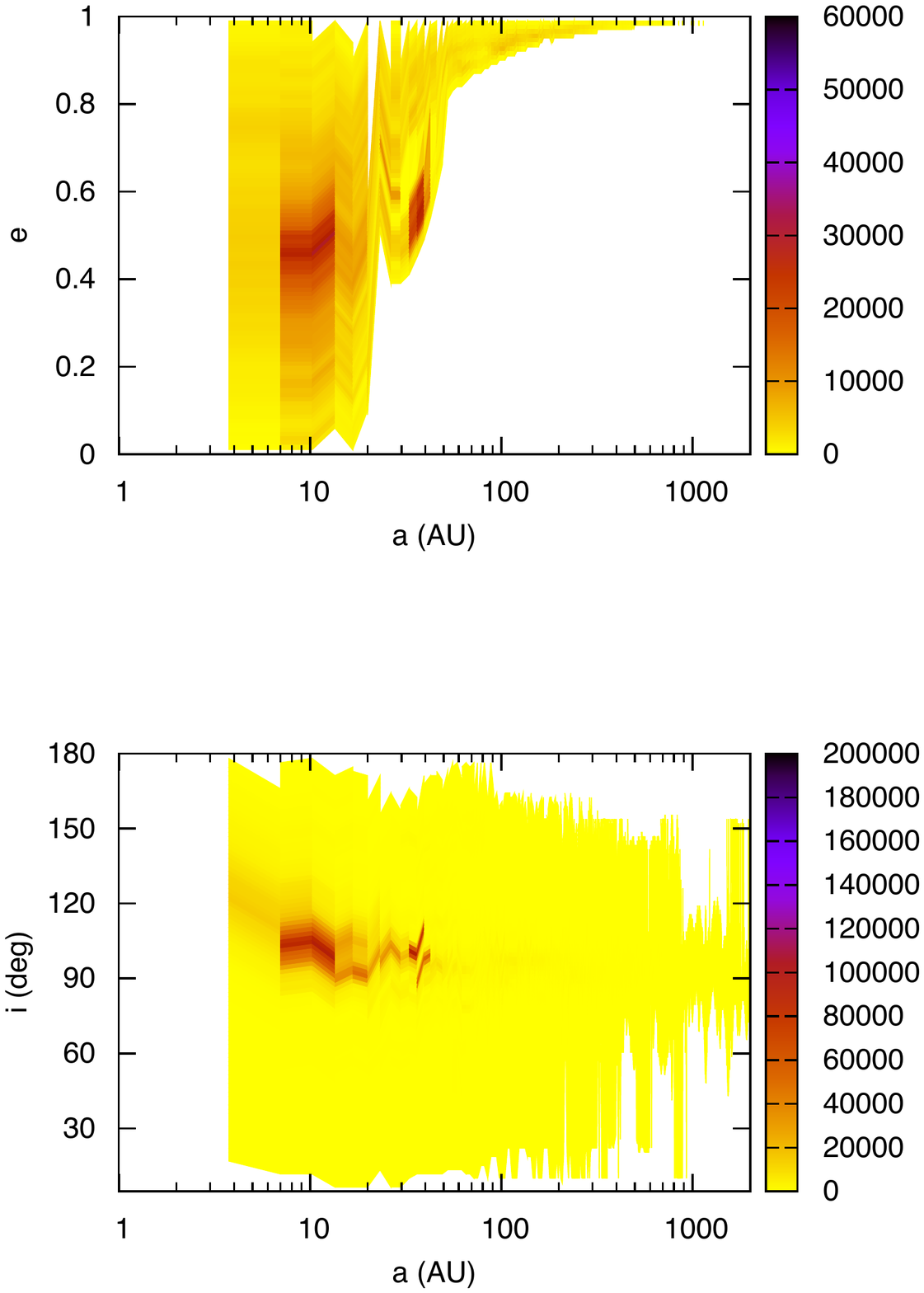}
  \includegraphics[width=0.49\linewidth]{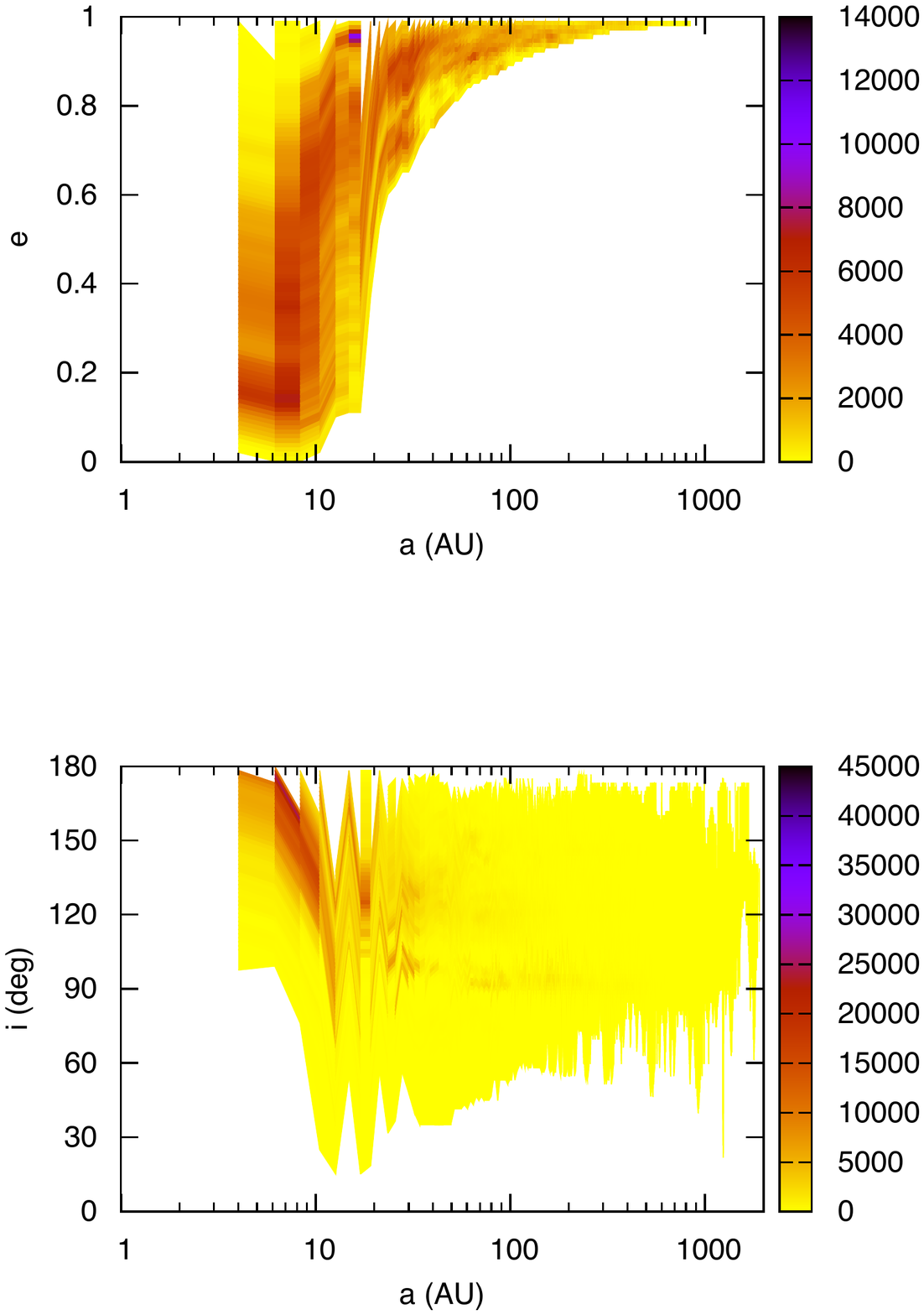}
  \caption{Residence maps of 2008 YB$_{3}$'s orbit clones (left panels)
  and 2005 VD's orbit clones (right panels). The units of
  the colorscales are $ 1.25\times 10^{-3} \times \,\mathrm{yr}^{-1}  \times \,\mathrm{AU}^{-1} 
 \times \deg^{-1} \times  \,\mathrm{Number of particles}^{-1}$ for the $a-i$ plot
  and $0.075 \times \,\mathrm{yr}^{-1} \times \,\mathrm{AU}^{-1} \times
\,\mathrm{Number of particles}^{-1}$ for the
  $a-e$ plots.}
  \label{fig:residence}
\end{figure*}

In Fig.~\ref{fig:residence} we plot the residence-time in $10^8$yr for both 2008 YB$_{3}$ and 2005 VD as a function of semi-major axis, $a$, and inclination, $i$, and also as a function of semi-major,  $a$, axis and eccentricity, $e$. This value is a representation of how probable it is to find an object at a certain orbit, represented by its orbital parameters, at a certain time during the integration. These plots were generated as follows: using the output of the numerical simulations of $10^8$ yr duration and sampling frequency $1/100 yr$, we counted the number of times that the orbit of each clone is within the bounds of a cell of dimensions, $1000 \,\mathrm{AU}/1500$ and $180^o/500$ for the $a-i$ plots and $1000 \,\mathrm{AU}/1500$ and $1/500$ for the $a-e$ plots.

For 2008 YB$_{3}$, the residence-time indicates that the inclination remained about its present values for most of its lifetime (see Fig.~\ref{fig:residence} left-hand panel). One of the most populated spots in the $a-e$ plot is located where the objects are found at present. The other one is located at semi-major axes with values between $20$AU and $30$AU, suggesting the TNb as a plausible region of provenance.

The picture given by the residence maps of 2005 VD is qualitatively different. From the $a-i$ plot, it is apparent a past as a retrograde object about the ecliptic. The $a-e$ plot shows a very populated spot at semi-major axis between $10$AU and $20$AU at very large, quasi-parabolic eccentricities. A ``handling down'' of the orbit towards a range of eccentricities closer to present values is apparent. Therefore, an immediate site of provenance from the inner Oort cloud is quite strongly indicated in these plots.

These diagrams give evidence of the apparent differences between 2005 VD and 2008 YB$_{3}$ in terms of dynamical evolution. From the comparison of the two panels on the right and the two panels on the left, it is apparent that 2005 VD has a longer residence-time in the planetary region, as well as in orbits with high inclinations. For 2008 YB$_{3}$, the inclination is always concentrated about its present values. But in the case of 2005 VD, the inclination concentrates approximately between $120\degr$ and $180\degr$. A picture of 2005 VD coming from quasi-parabolic orbits and 2008 YB$_{3}$ coming from the trans-Neptunian region emerges noticeably.

The $e$-fold decrease of the number of particles remaining in the integration as a function of time implies a similar dynamical lifetime for both objects of approximately $5.5$Myr. But in the case of 2008 YB$_{3}$, a larger residual population exists between $7$Myr and $20$Myr. As a consequence, we evaluate that the time spent by
2005 VD between the planets is shorter than in the case of 2008 YB$_{3}$.

To measure the efficiency of transporting the object to the inner Oort cloud from each of these two groups of orbits, we studied the number of particles remaining in the integration that reach aphelion distances greater than $100$AU. We noticed significative differences between $1$Myr and $2$Myr and between $4$Myr and $6$Myr at those time-intervals when the particles associated with 2005 VD beyond $100$AU exceed those of 2008 YB$_{3}$. Again, this suggests a site of provenance for 2005 VD at the Oort cloud and a site of provenance, for 2008 YB$_{3}$, at the trans-Neptunian region.

\section{Discusion and Conclusions}
\label{sec:conclusions}

We present new data of two retrograde objects, 2005 VD and 2008 YB$_{3}$. We show photometry of both objects and spectroscopy of 2008 YB$_{3}$.

We show lightcurves obtained for 2008 YB$_{3}$ from visible photometry and WISE data in the far-infrared. Both lightcurves are consistent, and from WISE data we can give an estimation of the rotational period of 16.388 $\pm$0.002 hr.

Colors of the two retrograde objects are similar, but there are some differences that could indicate a different surface composition. When compared with solar colors, 2008 YB$_{3}$ is moderately red while 2005 VD colors are more similar to the colors of the Sun. 

When compared with other minor bodies in the solar system, we find similarities between  the colors and spectra of 2008 YB$_{3}$  and those of the red Trojans (as defined in \citeauthor{eme11} \citeyear{eme11}), the nuclei of comets (either nuclei of short period comets or Damocloids), and the population of grey Centaurs as shown in
\cite{mellic12}. For 2008 YB$_{3}$, there is an estimation of the albedo in the visible of $\rho_{V}\sim $0.025 $\pm$ 0.004 (V. Al\'i-Lagoa, personal communication). This estimation, based on WISE observation, is compatible with other low albedo surfaces. With these new observations, we confirm that the ultra-red matter present on some Centaurs is missing from the surface of 2008 YB$_{3}$. All of this, suggests that the surface of this object lacks ice and is probably covered by a layer of silicates (amorphous pyroxenes and/or olivine).

We present colors of 2005 VD in the visible. They are affected by a larger error due to the faintness of this object but clearly show
that it is the bluest of all of the populations used in the comparison. Neutral colors in the outer solar system are typically associated with high-albedo objects covered by water ice \citep{pin08}. They are also typical of low-albedo surfaces of primitive minor objects, covered with carbons and silicates optically inactive at these wavelengths. Without a measurement of the albedo or observations in the NIR, it is not possible to distinguish between these two kinds of surfaces. However, based on the orbital characteristics of 2005 VD we favor the second option as the most probable composition of the surface of 2005 VD. 

The differences in color between 2005 VD and 2008 YB$_{3}$ are small but consistent with the results of the modeling of their dynamical evolution. We consider particularly interesting the comparison of the colors of both objects with the two groups of Centaurs that arise from the study of their colors. Trying to disentangle the bi-modality of
the color distribution of this population, \citet{mellic12} study links between their dynamical history and their surface physical properties. They find that the differences in colors could be related to different thermal processing of their surfaces. The objects in the grey lobe of the Centaurs have probably spent more time at short distances from the Sun and, as a result, they passed by more active epochs than the objects in the red lobe. Based on this, the authors proposed activity is probably the reason why the colors of the grey Centaurs are more similar to those of comet nuclei and other objects covered by mantles of dust. 

In our case, data of both, 2005 VD and 2008 YB$_{3}$, suggest a lack of volatiles on the surface and a composition similar to objects that have suffered from activity. The
small difference between them, with 2005 VD being more neutral than 2008 YB$_{3}$, suggests the first has a surface more processed than the latter. This is in agreement with the results of the dynamical studies. The residence times of both objects indicate that 2005 VD spends a larger lapse of time in the region of the giant planets, reaching shorter perihelion distances in the region of the near-earth asteroids, which is in agreement with a thicker and more neutral dust mantle lacking ice content.

We cannot reach a unique solution about the site of provenance of these objects.  Though retrograde orbits are only compatible with a common origin in the Oort cloud, the close evolution of these two particular objects presents some differences that noticeably suggest that 2005 VD is coming from quasi-parabolic orbits, while 2008 YB$_{3}$ is  coming from the trans-Neptunian region. In the case of  2008 YB$_{3}$, its evolution would be dominated by successive encounters with Jupiter and Saturn, resulting in the
insertion of the object from the TNb or a more distant reservoir at $Q\sim$70 AU into its present orbit, while the dynamics of 2005 VD is dominated by close encounters with planet Jupiter and a larger residence time in the planetary region.\\

 In summary:

\begin{itemize}
\item 2005 VD and 2008 YB$_{3}$ show colors and spectra (in the case of 2008 YB$_{3}$) similar to the colors and spectra of other populations of minor bodies likely covered by mantles of silicates and depleted of ices. 

\item 2005 VD is more neutral and appears on the blue end of the color-color plot, as shown in Fig.~\ref{fig:colors}, while 2008 YB$_{3}$, is reddish. This suggests that the surface of the former is more processed.

\item The study of integrations of their orbits shows that these objects have been retrograde for most of their histories. From our studies the injection of their orbits from the Oort cloud is probably different: 1) For 2008 YB$_{3}$,  the TNb as a plausible region of provenance is suggested;  2) For 2005 VD, an immediate site of provenance from the inner-Oort cloud is quite strongly indicated.

\item From the study of the surface characteristics and dynamical evolution of both objects, they have probably suffered from activity related to ices' volatilization that resulted in a surface covered by a mantle of silicates. The larger residence-time of 2005 VD in the planetary region of the giant planets with possible semi-major axis between 3 and 10 AU is a plausible explanation for the extreme color of this object in the visible.
\end{itemize}

\begin{acknowledgements}
 {\it NPA would like to thank the Instituto Nacional de Ci\^encia e Tecnologia de Astrof\'isica do Brasil, financed by CNPq and FAPESP, for its support when visiting as Specialized Researcher. MDM acknowledges contribution by ANPCyT through PICT 218/2007 and CONICET PIP 11220090100461/2010. JMC, DL, and PHH were supported by diverse grants and fellowships by CNPq, FAPERJ, and CAPES. E.C. acknowledges support by the Fondo Nacional de Investigaci\'on Cient\'ifica y Tecnol\'ogica (proyecto No. 1110100) and the Chilean Centro de Excelencia en Astrof\'isica y Tecnolog\'ias Afines (PFB 06). Part of this work is based on observations made with the Italian Telescopio Nazionale Galileo (TNG) operated on the island of La Palma by the Fundaci\'on Galileo Galilei of the INAF (Istituto Nazionale di Astrofisica) at the Spanish Observatorio del Roque de los Muchachos of the Instituto de Astrofisica de Canarias. This publication makes use of data products from the Wide-field Infrared Survey Explorer, which is a joint project of the University of California, Los Angeles, and the Jet Propulsion Laboratory/California Institute of Technology, funded by
the National Aeronautics and Space Administration.}

\end{acknowledgements}


\bibliographystyle{aa}
\bibliography{bibCentaurs}

\end{document}